\documentclass[amsmath,amssymb,12pt, eqsecnum]{revtex4}

\usepackage{hyperref}
\usepackage{multirow}
\usepackage{array}
\usepackage{sidecap}
\usepackage{subfigure}
\usepackage{graphicx}
\usepackage{epstopdf}
\usepackage{dcolumn}
\usepackage{bm}

\newcommand{\expec}[1]{\left\langle #1\right\rangle}
\newcommand{\half}[0]{\frac{1}{2}}
\newcommand{\pd}[2]{\frac{\partial #1}{\partial #2}}
\newcommand{\ld}[0]{\mathcal{L}}

\newcommand{\dd}[0]{\textrm{d}}
\newcommand{\qsubrm}[2]{{#1}_{\scriptsize{\textrm{#2}}}}

\newcommand{\ket}[1]{|#1\rangle}

\newcommand{\rbm}[1]{{\bf{#1}}}

\newcommand{\ci}[0]{\textrm{i}}

\newcommand{\defn}[0]{\equiv}

\newcolumntype{V}{>{\centering\arraybackslash} m{.4\linewidth} }

\newcommand{\fref}[1]{{Figure \ref{#1}}}

\def\be{\begin{equation}}
\def\ee{\end{equation}}
\def\bea{\begin{eqnarray}}
\def\eea{\end{eqnarray}}

\let\oldsqrt\sqrt
\def\sqrt{\mathpalette\DHLhksqrt}
\def\DHLhksqrt#1#2{%
\setbox0=\hbox{$#1\oldsqrt{#2\,}$}\dimen0=\ht0
\advance\dimen0-0.2\ht0
\setbox2=\hbox{\vrule height\ht0 depth -\dimen0}%
{\box0\lower0.4pt\box2}}

\begin{document}

\title{Charge, junctions and the scaling of domain wall networks}
\author{Richard A. Battye}
\email{richard.battye@manchester.ac.uk}
\affiliation{Jodrell Bank Centre for Astrophysics, School of Physics and Astronomy, The University of Manchester, Manchester M13 9PL, U.K}
\author{Jonathan A. Pearson}
\email{jp@jb.man.ac.uk}
\affiliation{Jodrell Bank Centre for Astrophysics, School of Physics and Astronomy, The University of Manchester, Manchester M13 9PL, U.K}

\date{\today}
\begin{abstract}
It has been shown that superconducting domain walls in a model with $U(1) \times \mathbb{Z}_2$ symmetry can form long-lived loops called kinky vortons from random initial conditions in the broken field and a uniform charged background in $(2+1)$ dimensions. In this paper we investigate a similar model with a hyper-cubic symmetry coupled to an unbroken $U(1)$ in which the domain walls can form junctions and hence a lattice. We call this model the charge-coupled cubic-anisotropy (CCCA) model. First, we present a detailed parametric study of the $U(1)\times \mathbb{Z}_2$ model; features which we vary include the nature of the initial conditions and the coupling constants. This allows us to identify interesting parameters to vary in the more complicated, and hence more computationally intensive, CCCA models. In particular we find that the coefficient of the interaction term can be used to engineer three separate regimes: phase mixing, condensation and phase separation with the condensation regime corresponding to a single value of the coupling constant defined by the determinant of the quartic interaction terms being zero. We then identify the condensation regime in the CCCA model and show that, in this regime, the number of domain walls does not scale in the standard way if the initial conditions have a sufficiently high background charge. Instead of forming loops of domain wall, we find that, within the constraints of dynamic range, the network appears to be moving toward a glass-like configuration. We find that the results are independent of the dimension of the hyper-cube.
\end{abstract}

 \maketitle

\section{Introduction}

One idea to explain the mysterious dark energy which appears to pervade the Universe is that it is due to a network of domain walls which has frozen into some kind of static configuration which is akin to a soap film~\cite{bucher_spergel_1999, battye_bucher_spergel_1999}. Such a model predicts an equation of state with $w=P/\rho=-2/3$ and can be represented in cosmological perturbation theory by an elastic medium with rigidity and a relativistic sound speed (see ref.~\cite{BM-2007} for a detailed discussion). An important question is whether such a static network can be created from random initial conditions. This would require the formation of either a regular lattice or a more amorphous glass-like state.

In this paper we aim to develop further the idea that charge condensation onto the walls can be used to slow down and possibly even prevent the natural propensity of random wall configurations to collapse under tension, lose energy as fast as causality allows and scale in self-similar way with the number of walls being $\propto t^{-1}$. This might be possible in models where the domain wall forming fields are coupled to a complex scalar field whose potential has an unbroken $U(1)$ symmetry. 

A model with $U(1)\times \mathbb{Z}_2$ symmetry was discussed  in ref.~\cite{BattyePearson-kvform} in (2+1) dimensions. It was shown that long-lived loop configurations can be formed from initial conditions where the domain wall forming field is distributed randomly among the two vauca and the $U(1)$ symmetric field is set to a uniform charge background. These loop configurations, known as kinky vortons~\cite{BattyeSut-KV,BS-KVEos}, are stabilized by the Noether current and associated charge in a similar way to cosmic vortons~\cite{davis_shellard_1988_1}. 

The $U(1)\times \mathbb{Z}_2$ model does not allow for junctions between domain walls and one might think that they are necessary in order for the model to form a lattice or glass-like state. Junctions are present in the cubic anistropy model which is a model for an $N$-component field whose potential has minima on vertices or the centre of the faces of a hyper-cube. The dynamics of this model have been be shown to lead to something close to the standard scaling regime within the narrow dynamical range possible in numerical simulations~\cite{battye_moss_o3}.

We will start by performing a detailed parametric study of the $U(1)\times \mathbb{Z}_2$ model. We will show that our previous results are largely independent of the way the initial conditions are set up and will investigate the scaling behavior as a function of the coupling constants. Importantly we find that the specific choice of the coefficient of the cross-coupling used in ref.~\cite{BattyePearson-kvform} is a boundary between two regimes:  phase mixing which takes place for lower values and phase separation for higher values. Only at the exact value used in ref.~\cite{BattyePearson-kvform} does the phase condensation, necessary for the modified scaling behaviour observed, take place. We then go on to study a cubic anisotropy model coupled to a complex scalar field. We will call this model the charge-coupled cubic anisotropy  (CCCA) model. Once we have identified the charge condensation regime, we find that this model behaves a similar way to the $U(1)\times \mathbb{Z}_2$ model with charge becoming localized on the walls and we go on to discuss the nature of the configurations which are produced.

\section{Models and Numerical Implementation}
\label{sec:model_intro}
The models we will investigate have two interacting fields $\Phi(x^{\mu})$ and $\sigma(x^{\mu})$ such that the equations of motion permit domain wall solutions in the $\Phi$-field and $\sigma$ has a conserved Noether charge. We will consider models with a Lagrangian density which has the generic form
\bea
\label{eq:sec2:full-pot}
\ld = \half \partial^{\mu}\Phi \partial_{\mu}\Phi + \half \partial^{\mu}\sigma\partial_{\mu}\bar{\sigma} - V_1\left(\Phi\right) - V_2\left(|\sigma|\right) - \beta |\Phi|^2|\sigma|^2,
\eea
where the conserved charge of the complex scalar field $\sigma(x^{\mu}) \in \mathbb{C}$ is given by
\bea
Q = \int \dd^2x\, \qsubrm{\rho}{Q}=\frac{1}{2\ci}\int \dd^2x\,\left( \dot{\sigma}\bar{\sigma} - \dot{\bar{\sigma}}\sigma\right),
\eea
and $V_1(\Phi)$ has discrete minima. We will take the specific form 
\bea
V_2(|\sigma|) = \frac{\lambda_{\sigma}}{4}\left( |\sigma|^2 - \eta_{\sigma}^2 \right)^2,
\eea
and the symmetry breaking term $V_1(\Phi)$ to be one of two cases
\begin{subequations}
\bea
\label{eq:pot_1a_kv}
V_{1}^{(a)}(\Phi) &=& \frac{\lambda_{\Phi}}{4}\left( \Phi^2 - \eta_{\Phi}^2\right)^2,\quad \Phi(x^{\mu}) \in \mathbb{R};\\
\label{eq:pot_1b_mca}
V_{1}^{(b)}(\Phi) &=& \frac{\lambda_{\Phi}}{4}\left( |\Phi|^2 - \eta_{\Phi}^2\right)^2 + \epsilon \sum_{i=1}^N\phi_i^4,\quad \Phi = \left(\phi_1, \phi_2, \ldots, \phi_N\right), \phi_i(x^{\mu}) \in \mathbb{R}.
\eea
\end{subequations}
We can rescale the energy and length units allowing us, without loss of generality,  to define $\lambda_{\Phi} = \eta_{\Phi} = 1$. The model defined by  $V_{1}^{(a)}$ is a special case of $V_{1}^{(b)}$ with $\epsilon = 0, N=1$  and taking $\beta = 0$ it becomes the non-interacting discrete $\mathbb{Z}_2$ model which has a static kink (domain wall) solution. For $\beta \neq 0$, the model has superconducting domain wall solutions~\cite{hodges_1988,BattyeSut-KV} for specific values of the parameters. The model defined by  $V_{1}^{(b)}$ for $\epsilon = 0$ has a global $O(N)\times U(1)$ symmetry. For $\beta=0$ the $O(N)$ symmetry is broken to a discrete hyper-cubic symmetry group when $\epsilon\ne 0$; for $\epsilon <0$ the points of the vacuum manifold are the $2N$ faces of an $N$-dim hypercube,  whereas for $\epsilon >0$ it is the vertices of the hypercube. For $\epsilon>0$ and $N >2$ there are walls of a number of different tensions whereas only 2 different tensions exist for any $N$ with $\epsilon <0$. 

The vacuum in the $\epsilon >0$ case in the presence of charge needs subtle consideration as we shall now describe. We shall continue with an $N=2$ case for simplicity, but the method can be generalized by the embedding $\mathbb{R}^N \rightarrow S^{N-1}$. First let us write $\phi_1 = |\Phi|\cos\theta, \phi_2 = |\Phi|\sin\theta$ so that the $N=2$ CCCA model's potential reads
\begin{eqnarray}
\label{eq:sec2:modccccccca}
V = \frac{\lambda_{\Phi}}{4}\left( |\Phi|^2 - \eta_{\Phi}^2\right)^2 + \frac{\lambda_{\sigma}}{4}\left( |\sigma|^2 - \eta_{\sigma}^2 \right)^2+\beta|\Phi|^2|\sigma|^2+ \epsilon|\Phi|^4 \left( \cos^4\theta + \sin^4\theta\right).
\end{eqnarray}
As all terms are positive definite (except the last term) and the (uncharged) vacuum has
\begin{subequations}
\begin{eqnarray}
|\Phi|^2 = \frac{\lambda_{\Phi}\eta_{\Phi}^2}{2\epsilon + \lambda_{\Phi}},&\qquad& \epsilon >0,\\
|\Phi|^2 = \frac{\lambda_{\Phi}\eta_{\Phi}^2}{4\epsilon + \lambda_{\Phi}},&\qquad& \epsilon <0,
\end{eqnarray}
\end{subequations}
the final term in (\ref{eq:sec2:modccccccca}) is only one which can select the discrete points that make up the vacuum manifold. It is now a trivial task to note that the vacuum for the $\epsilon >0$ case is constructed from the phase-set $\theta_> \defn \{\pi/4, 3\pi/4, 5\pi/4, 7\pi/4\}$ and in the $\epsilon <0$ case by the phase-set $\theta_< \defn \{0,\pi/2, \pi, 3\pi/2 \}$ (which represent the vertices and faces respectively of a square). The height of a potential barrier between adjacent minima can then be computed, $\Delta V \defn V(\theta \in \theta_<) - V(\theta \in \theta_>)$, and is easily found to be $\Delta V = \half|\Phi|^4\epsilon$. The strong dependence of the height of the potential barrier $\Delta V$ upon $|\Phi|$ means that if $|\Phi|/\eta_{\Phi}<1$ only a very small perturbation is needed to surmount the potential barrier and move between minima. In the presence of coupling to charge (i.e. $\beta \neq 0$) this effect is so acute that regions of space occupying the discrete minima do not form in the $\epsilon >0$ case which has $|\Phi|/\eta_{\Phi}<1$ in the vacuum (whereas $|\Phi|/\eta_{\Phi}>1$ in the vacuum in the $\epsilon <0$ case) and hence in all results presented in this paper we use $\epsilon <0$.

In order for the discrete hyper-cubic symmetry to be broken locally one requires that the effective mass for $\Phi$ is negative, $\qsubrm{m}{eff,$\Phi$}^2<0$. Including the interaction term corresponds to the constraint
\bea
 \qsubrm{m}{eff,$\Phi$}^2=\beta |\sigma|^2 - \frac{\lambda_{\Phi}}{2}\eta_{\Phi}^2<0,
\eea
and hence one requires for symmetry breaking that $|\sigma|^2<\kappa$ where 
\bea
\kappa \defn\frac{\lambda_{\Phi}\eta_{\Phi}^2}{2\beta}.
\eea
The fact that $|\sigma|^2$ varies from place to place means that the symmetry breaking condition $\qsubrm{m}{eff,$\Phi$}^2<0$ is not necessarily met locally providing a rich landscape of phenomenologies.

For our numerical work we evolve the equations of motion in Minkowski spacetime in (2+1)-dimensions by discretizing on a regular square grid of $P$ grid-points along each side with grid-spacing $\Delta x$. We approximate spatial derivatives to fourth order and time evolution is achieved using a second-order leapfrog algorithm with time-step size $\Delta t$. We always use $\Delta x = 0.5$, $\Delta t = 0.1$ and we use both $P=1024$ and $P=4096$ -- the former for computational brevity and the latter for high dynamical range. The boundary conditions we employ are periodic which sets an upper bound, $\tau$, on the time a given simulation can be used to deduce dynamical information. This is the time taken for an emitted signal to interact with itself after having crossed the periodic boundaries - the so called ``light crossing time''. It is simple to show that $\tau = P\Delta x/2\Delta t$. Our results on the evolution of the scaling of domain walls are only valid upto this time; however, structures that persist after this time are of interest since they indicate stability of the wall network against perturbations due to radiation which is ubiquitous within the simulation.

Typically we use initial conditions that could describe a system directly after a phase transition in a uniform charged background. To achieve this, we assign every grid-point a random position on the vacuum manifold (with $\dot{\Phi} = 0$), and set $\sigma = Ae^{-\ci\omega t}$. This corresponds to an initially homogeneous charge density $\qsubrm{\rho}{Q}(0) = A^2\omega$ specified by two parameters $A$, $\omega$. An unphysical side-effect of our initial conditions is that large energy gradients are setup between adjacent grid-points. To remedy this we introduce a damping term which acts on only the  $\Phi$-field for the first 200 time-steps in our simulation -- this reduces the overall energy, aiding condensation into domains, whilst leaving the total charge constant. The dynamic range of a simulation is $200 < t < \tau$: the time period over which physically meaningful results can be extracted.

In order to investigate the scaling dynamics of the domain walls, we use an algorithm to compute the number of domain walls at each timestep. This works by deciding if adjacent grid-points occupy the same point on the vacuum manifold -- if they do, then no wall exists, and if different the wall count is incremented. We compute scaling exponents by fitting a power-law to the number of domain walls, $\qsubrm{N}{dw} \propto t^{-\gamma}$, in bins of 200 timesteps over the dynamic range. 

\section{Parametric Study of the $U(1)\times \mathbb{Z}_2$ Model}
\label{sec:kinky-vorton}
The aim of this section is to present results from a parametric investigation of the $U(1)\times \mathbb{Z}_2$ model  using (\ref{eq:pot_1a_kv}) as the symmetry breaking potential. These results are an extension to those presented in our previous paper~\cite{BattyePearson-kvform} allowing us to identify interesting effects in the computationally less expensive $U(1)\times \mathbb{Z}_2$ model which might be important in the CCCA model. Typically we will vary one parameter and keep the others fixed to fiducial values $A=1$, $\omega=1$, $\lambda_{\sigma}=1$, $\beta=1/2$ and $\eta_{\sigma}=1$.

\subsection{Variation of initial conditions and $\eta_{\sigma}$}

\begin{figure*}[!t]
      \begin{center}
	{\includegraphics[scale=1.5]{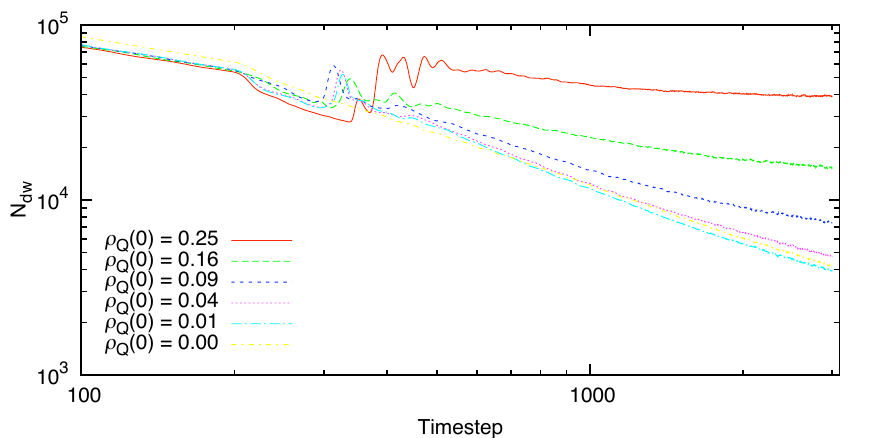}}
      \end{center}
\caption{Evolution of the number of walls for various $\qsubrm{\rho}{Q}(0)$, where we have fixed $\omega = 1$, and varied $A$. Each plot is an ensemble average over 10 realisations with $P = 1024$. In all plots we use $\beta = 1/2$, $\eta^2_{\sigma}=3/4$. Also shown for comparison is the $\qsubrm{N}{dw} \propto t^{-1}$ scaling law adhered to by the $\qsubrm{\rho}{Q}(0) = 0.0$ system.}\label{fig:sec3-kvvaryalpha}
\end{figure*}  

\begin{figure*}[!t]
      \begin{center}
\includegraphics[scale=1.3]{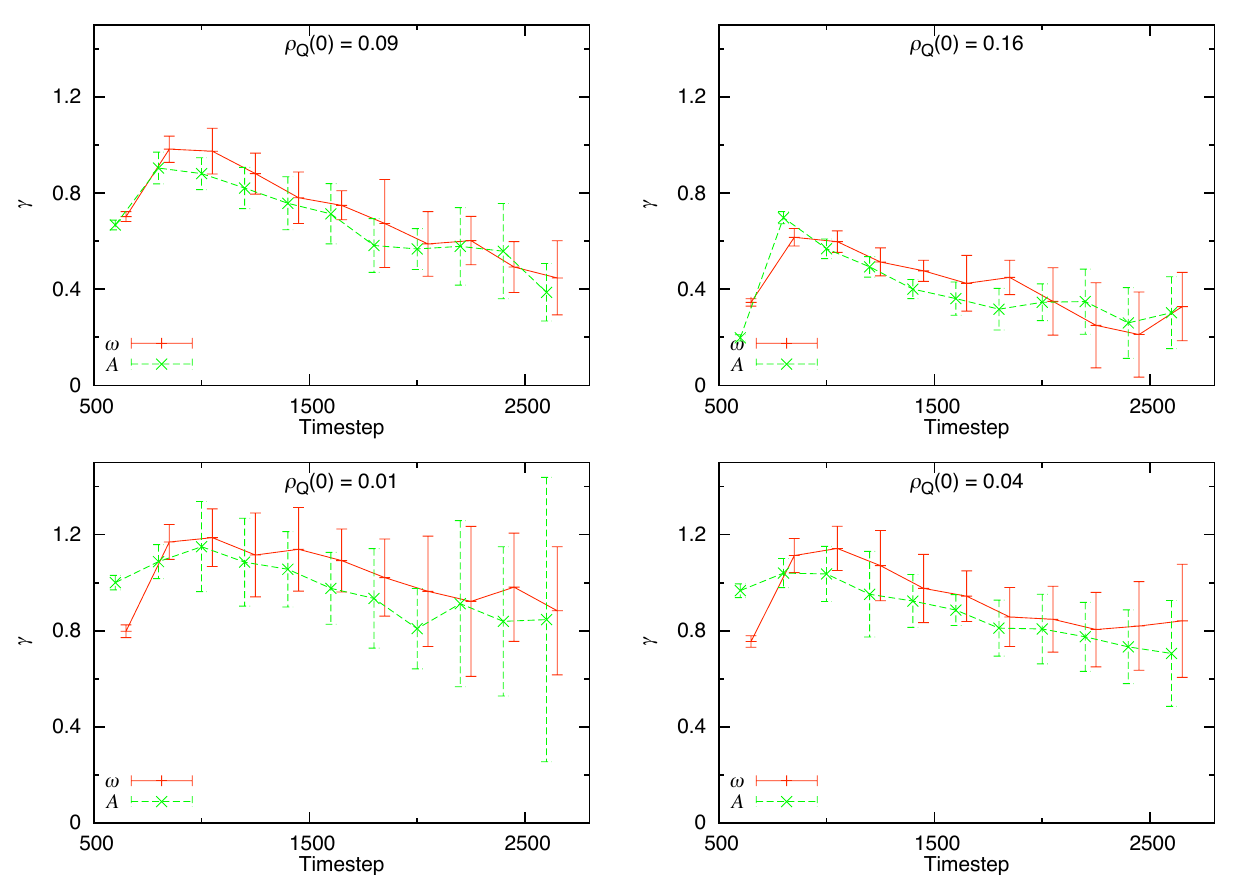}
      \end{center}
\caption{Scaling exponents, where $\qsubrm{N}{dw} \propto t^{-\gamma}$, for different values of $\qsubrm{\rho}{Q}(0)$ created by varying either $A$ or $\omega$, whilst keeping the other fixed. Each panel is for a specific value of $\qsubrm{\rho}{Q}(0)$, where the varied parameter is shown in the legend. All plots where $\omega$ is varied are offset by 50 time-steps to allow the $1\sigma$-error bars to be seen. We have used $P = 1024, \eta_{\sigma}^2 = 3/4$ over 10 ensembles.}\label{fig:sec3:kv-scaleesxps-varymethod}
\end{figure*}  

We have performed a set of simulations varying the initial charge parameters $A$ and $\omega$. 
\fref{fig:sec3-kvvaryalpha} shows the evolution of the number of walls as a function of varying initial charge density -- where we have used $A$ to vary $\qsubrm{\rho}{Q}(0)$. These are equivalent to results presented in ref.~\cite{BattyePearson-kvform}. As one can clearly see the evolution diverges from the standard scaling law as $\qsubrm{\rho}{Q}(0)$ increases; a conclusion which is further strengthened by considering the scaling exponents presented in \fref{fig:sec3:kv-scaleesxps-varymethod}. Also shown in \fref{fig:sec3:kv-scaleesxps-varymethod} are scaling exponents for simulations where $A$ and $\omega$ are separately varied to give the same values of $\qsubrm{\rho}{Q}(0)$. It is clear that, within the computed errors, the two methods give similar results; albeit those with varying $\omega$ being marginally lower.

\begin{figure*}[!t]
      \begin{center}
\subfigure[\,Number of walls.]{\includegraphics[scale=1.2]{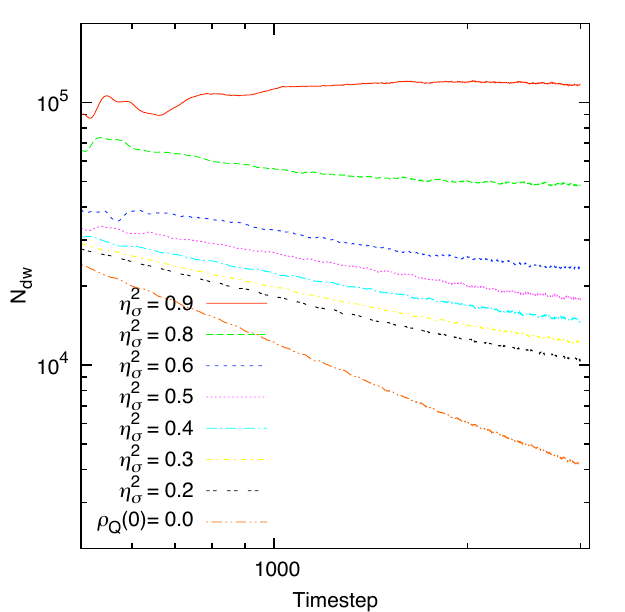}}\quad
\subfigure[\,Scaling exponent]{\includegraphics[scale=1.2]{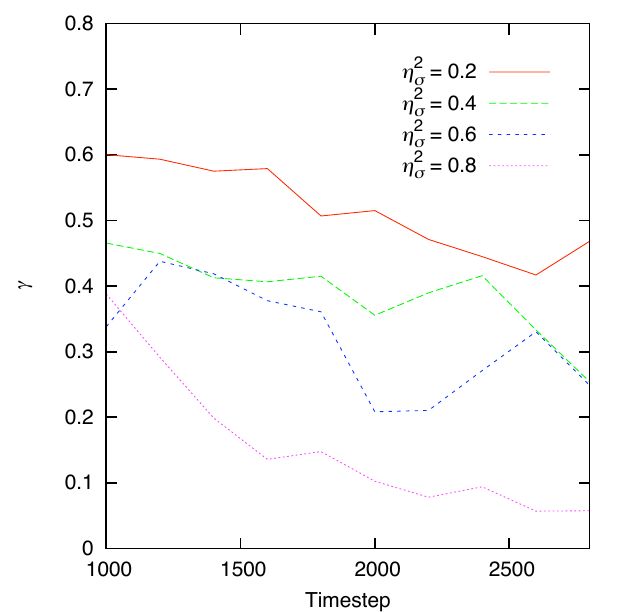}}
      \end{center}
\caption{Evolution of the number of walls and scaling exponents $\qsubrm{N}{dw} \propto t^{-\gamma}$ for various $\eta_{\sigma}^2$ with $P=1024$. Each case is an ensemble average over 10 realisations. In all plots we use $A = 1/2$, $\omega = 1$, so that $\qsubrm{\rho}{Q}(0) = 0.25$ and $\beta = 1/2$. We have only presented a few of the scaling exponents of $\eta_{\sigma}^2$, for clarity -- the trend is clear. }\label{fig:sec3:kv-varyeta}
\end{figure*}

In \fref{fig:sec3:kv-varyeta} we present the evolution of the number of walls as we vary the parameter $\eta_{\sigma}^2$ for $\qsubrm{\rho}{Q}(0) = 0.25$. Evidently, we see a  deviation from scaling for the range of $\eta_{\sigma}^2$ presented. As $\eta_{\sigma}^2$ is increased the scaling exponent $\gamma$ diverges from the scaling value $\gamma = 1$. The known kinky vorton  solutions exist only for $\eta_{\sigma}^2 > 1/2$ since the analytic solution on which they are based requires this constraint, and therefore it is interesting to see that it appears that a superconducting wall solution exists outside the analytically known range.

\subsection{Variation of $\beta$ and phase separation}

The parameter whose variation produces the most interesting results is the cross-coupling constant $\beta$ since by varying its value we can achieve three different kinds of behavior which we shall call phase mixing, condensation and separation. In order to understand this behaviour we have done three things. First we will present results from 1D simulations which make it easy to see what is going on, some 2D simulations equivalent to those presented in the previous section and finally some simple analytic arguments which suggest the existence of the phase separation regime.

\begin{figure*}[!h]
      \begin{center}
	{\includegraphics[scale=0.85]{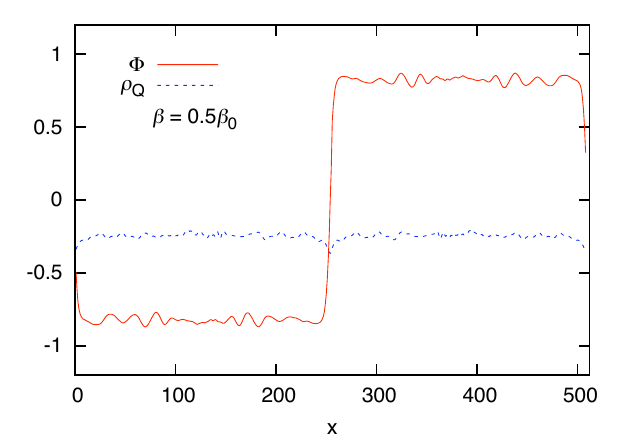}}
	{\includegraphics[scale=0.85]{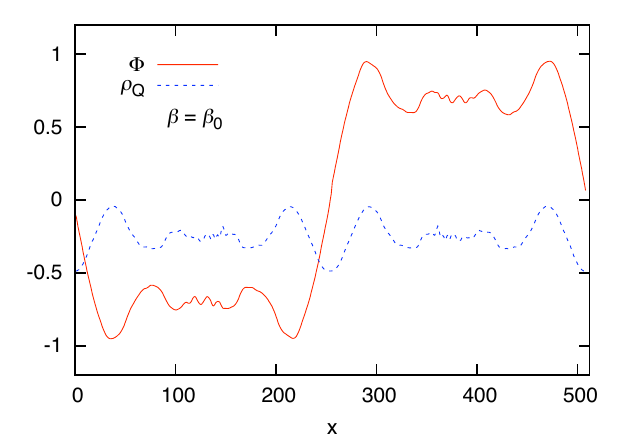}}
	{\includegraphics[scale=0.85]{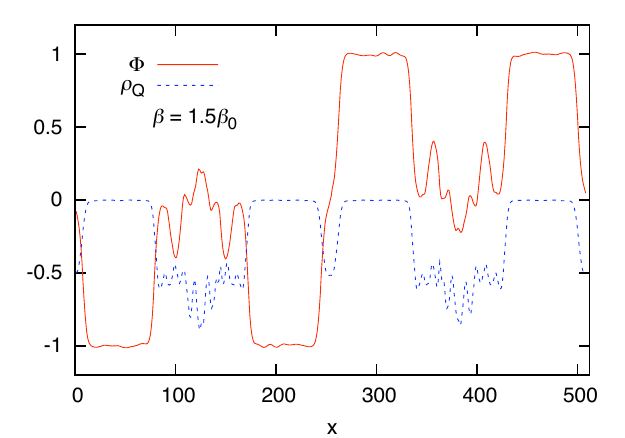}}
     \end{center}
\caption{Snapshot of the $\Phi$-field and charge density $\qsubrm{\rho}{Q}$ at $t=800$. The initial configuration has each half of the grid in each of the vacua with an initially homogeneous charge density with $\omega = 1$, $A=1/2$ (i.e. $\qsubrm{\rho}{Q}(0) = 0.25$). As indicated, we display the results from $\beta = \half\beta_0,\beta_0,\frac{3}{2}\beta_0$, where $\beta_0 = \half$. For $\beta < \beta_0$, one observes the homogeneous charge distribution and a domain wall only between vacuum states. For $\beta = \beta_0$, the charge distribution has evolved to localize on the domain wall separating vacuum states. For $\beta > \beta_0$, there is the production of a domain wall between the components of the mixture as well as walls between vacuum states; thus, domain walls interpolating between vacua $\Phi =1 $ and $\Phi=-1$, and between $\Phi =0$ and $|\Phi|=1$.}\label{fig:sec3-kv-phase_sep-1d}
\end{figure*}

For our 1D simulations the initial conditions have half of the grid with the $\Phi$-field in one vacuum and the other half in the other vacuum  and a homogeneous charged background. The state of the system is presented at $t=800$ in \fref{fig:sec3-kv-phase_sep-1d} for three different values of $\beta$. For reasons that will become clear later, we define $\beta_0 \defn \half\sqrt{\lambda_{\Phi}\lambda_{\sigma}} = \half$ and we will consider the cases $\beta = \half \beta_0, \beta = \beta_0$ and $\beta = \frac{3}{2}\beta_0$.

The case with $\beta = \half \beta_0$ has evolved to the standard kink solution and the charge density  is homogeneously distributed. The case with $\beta = \beta_0$ has evolved to a superconducting kink solution with the charge density localized on the kink between adjacent vacuum states. Away from the kink the charge density is lower but non-zero which leads to the $\Phi$-field being away from the vacuum points $\pm 1$. The final case, $\beta = \frac{3}{2}\beta_0$, leads to the production of two types of walls, one which is the superconducting wall and a second ``domain wall'' which interpolates between regions with and without charge -- the so-called phase separation regime. We call the first regime ($\beta <  \beta_0$) the \textit{phase mixing regime}, the second ($\beta = \beta_0$) the \textit{condensate regime} and the third ($\beta > \beta_0$) the \textit{phase separation regime}. Thus, we have a description of the system as being a mixture of two components: $\Phi$ and $\sigma$, and we can see that there are domain walls between states $\ket{\expec{\Phi} =0, \expec{\sigma} \neq 0}$ and $\ket{\expec{\Phi}\neq 0, \expec{\sigma} = 0}$, as well as between adjacent vacuum states $\ket{\Phi =+1}$ and $\ket{\Phi=-1}$. From here and on we shall call the wall between adjacent vacuum states a \textit{vacuum-wall} and the wall between the components a \textit{component-wall}. The vacuum-wall may also have a superconducting condensate localized upon it (as is the case in the $\beta = \beta_0$ system) in which case it is a superconducting vacuum-wall.

\begin{figure*}[!ht]
      \begin{center}
{\includegraphics[scale=0.9]{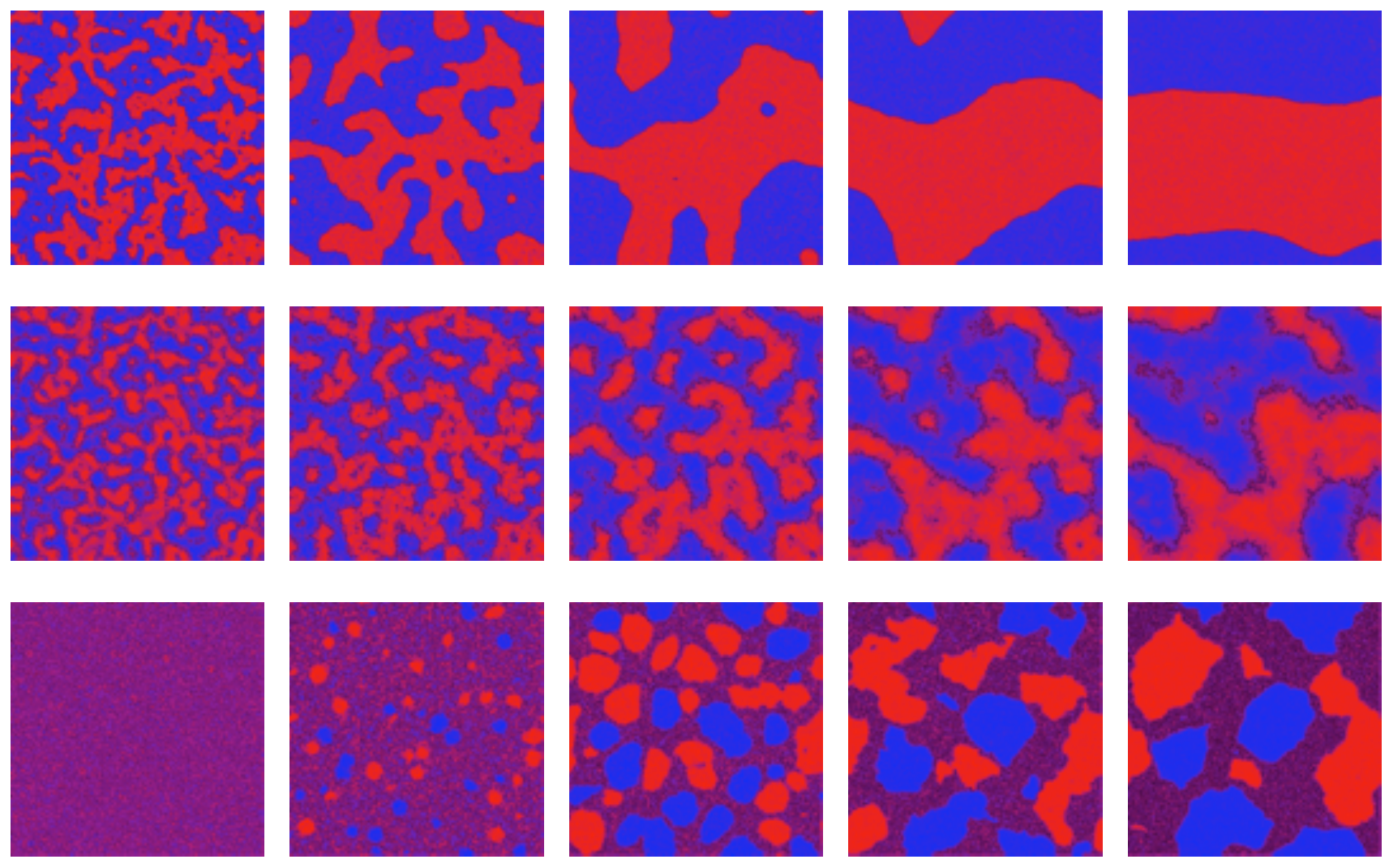}}
      \end{center}
\caption{Evolution of the $\Phi$-field for $\beta = \half\beta_0,\beta_0, \frac{3}{2}\beta_0$ (top to bottom) at times $t = 400, 800, 1600, 3200, 6400$ (left to right) with $\qsubrm{\rho}{Q}(0) = 0.25 ,\eta_{\sigma}^2 = 3/4$ and $P=1024$. Each colour/shade represents each of the two points on the vacuum manifold (with purple/darkest being $|\Phi|\approx 0$). The behavior observed in these three cases is similar to that presented in 1D in \fref{fig:sec3-kv-phase_sep-1d}. Vacuum-walls are observed when regions in different vacuum states are adjacent and component-walls where regions of vacuum occupation are adjacent to regions of extended charge density.}\label{fig:sec3-kv-phasesep}
\end{figure*}

We have evolved the 2D equations of motion using $\beta  = \half\beta_0,\beta_0, \frac{3}{2}\beta_0$ and we present results in \fref{fig:sec3-kv-phasesep}. We show the evolution of the $\Phi$-field from random initial conditions and a homogeneous background charge. We have fixed $\eta_{\sigma}^2 = 3/4, \qsubrm{\rho}{Q}(0) = 0.25$. We  see that the three regimes identified in 1D realize themselves in 2D with the production of vacuum-walls and component-walls where appropriate. The evolution of the $\beta = \half\beta_0$ case proceeds unimpeded by the charge which is confirmed by the computed scaling exponent $\gamma\approx 1$. As $\beta$ is increased towards $\beta_0$, the scaling exponent $\gamma\rightarrow 0$. For the case of $\beta = \beta_0$ there are vacuum-walls which clearly do not collapse in the standard way -- upon inspection of the other fields, we find high charge and current densities along the domain walls as explained in ref.~\cite{BattyePearson-kvform}. We have argued that the superconducting condensate has become naturally attracted to the vacuum-walls and the conserved current provides a restorative force which opposes the natural tension in the vacuum-walls.

The third case, $\beta = \frac{3}{2}\beta_0$, shows a very different picture. One can clearly see that regions where $\Phi$ occupies the discrete vacuum states (i.e. where $\Phi = \pm 1$) appear from a background of $|\Phi|\approx 0$. Upon inspection of the charge density, we find that $|\qsubrm{\rho}{Q}| \approx 0$ inside a region occupying a vacuum state and $|\qsubrm{\rho}{Q}|$ peaks in the regions where $|\Phi|\approx 0$. These observations corroborate those of the simple 1D system (\fref{fig:sec3-kv-phase_sep-1d}) and confirm the veracity of our classifications of phase mixing and separation. We rely on the visual observation that the domain wall network has not entered the standard scaling dynamics in the phase separation regime since the algorithm for calculation of $\qsubrm{N}{dw}$ is no longer valid. If we let $|\beta - \beta_0|\rightarrow 0$ in the phase separation regime (i.e. approach from above) then separation of the components still occurs but the extent of separation decreases as $|\beta - \beta_0|\rightarrow 0$.

One can understand the behavior observed in the simulations using simple analytic arguments. We will show that a component-wall exists between the $\Phi$ and $\sigma$ fields, for $\beta > \half \lambda_{\Phi}$ in the $U(1)\times \mathbb{Z}_2$ model. In particular, we substitute the ansatz for a $y$-directed vorton
\bea
\label{eq:sec3:kvanstaz-1}
\binom{\Phi}{\sigma} = \rho \binom{\cos(\theta/2)}{\sin(\theta/2)e^{\ci(\omega t + ky)}},
\eea
into the energy density where $\theta = \theta(x)$ and $\rho = \rho(x)$ is the total field density $\rho^2 = \Phi^2 + |\sigma|^2$. Assuming $\lambda_{\Phi} = \lambda_{\sigma}\defn \lambda$ we find that we can write the free energy
\bea
\label{eq:sec3:kvanstaz-2}
F = E-\mu Q = \int \dd^2x\, \left \{\frac{1}{8}\rho^2|\nabla\theta|^2 +{\Delta}\rho^4\sin^2\theta + \half|\nabla\rho|^2 + \frac{\lambda}{4} (\rho^2 - \eta_{\Phi}^2)^2 + \frac{\lambda}{4}\eta_{\sigma}^4 \right\},
\eea
where the charge and chemical potential are defined to be
\bea
Q \defn \omega \int \dd^2x\,\rho^2\sin^2(\theta/2),\qquad \mu \defn \frac{1}{2\omega}\left[{\lambda}\left( \eta_{\Phi}^2 - \eta_{\sigma}^2\right)+  \omega^2  + k^2\right],
\eea
and ${\Delta} \defn \frac{1}{4}(\beta -\half\lambda)$. 

For $\lambda \gg |\Delta|$, $\rho$ will be approximately constant and hence
\bea
F \approx\int \dd^2x\, \left\{ \frac{1}{8}\rho^2|\nabla\theta|^2 +{\Delta}\rho^4\sin^2\theta\right\}.
\eea
This is the double sine-Gordon model and is known to have solutions for sufficiently high $Q$ if $\Delta >0$ \cite{ivanov_1977}. The solutions have $\theta(x) = \pi$ for $|x| \ll R$ and $\theta(x) = 0$ for $|x| \gg R$, which is a region of charge with a domain wall separating the two phases. We note that $\Delta =0$ corresponds to $\beta =\beta_0= \half\lambda$ and hence this phase separation regime is valid for $\beta > \beta_0$.

\subsection{Zero Net Charge}
\begin{figure*}[!ht]
      \begin{center}
{\includegraphics[scale=0.4]{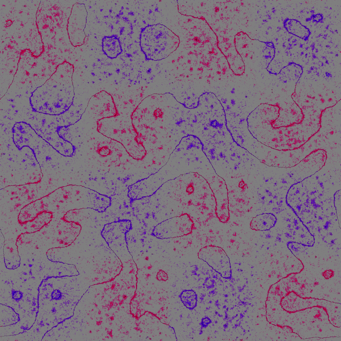}}\quad
{\includegraphics[scale=0.4]{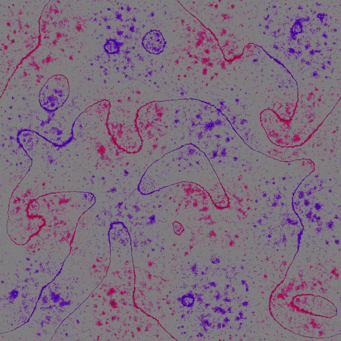}}\quad
 {\includegraphics[scale=0.4]{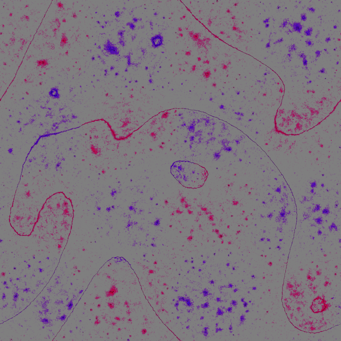}}\\\vspace{0.5cm}
{\includegraphics[scale=0.4]{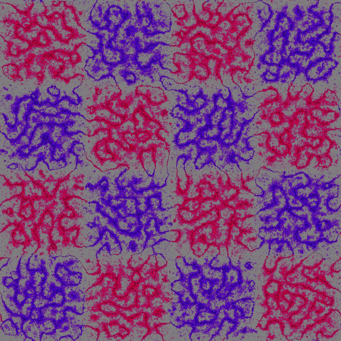}}\quad
{\includegraphics[scale=0.4]{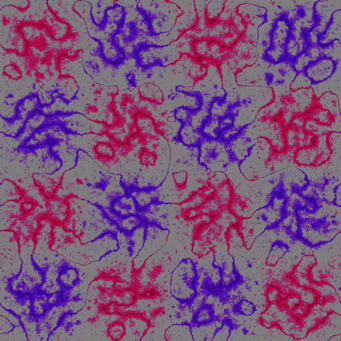}}\quad
{\includegraphics[scale=0.4]{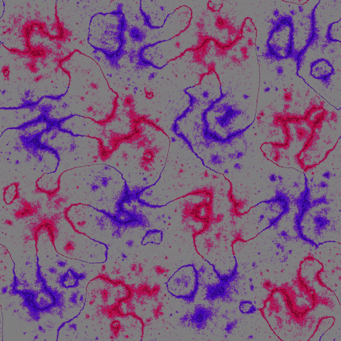}}
      \end{center}
\caption{Evolution of the charge density $\qsubrm{\rho}{Q}$ in simulations of the $U(1) \times \mathbb{Z}_2$ model in the condensate regime. Initial conditions have charge density $\qsubrm{\rho}{Q}(0) = 0.09, 0.25$ (top to bottom) in blocks as described in the main text. Images are at $t = 3200, 6400, 12800$ (left to right) and are coloured such that red/blue denotes positive/negative and grey where the value is less than $10\%$ of the maximum. We have used $P = 4096$ and  $\eta_{\sigma}^2 = 3/4$.}\label{fig:sec3:4096_kv_ccor_rho}
\end{figure*}

All the simulations we have presented so far have an overall net charge within the simulation box. Although it might be possible for some overall net charge to exist in the Universe, a more conservative idea would be that there is a zero net charge with regions of positive and negative charge having some correlation length different to that of the vacuum-wall forming field. In order to see if the phenomena observed in the case of a homogeneous charge persist in this case we have completed a set of simulations with $P = 4096$ and an initial charge-correlation length which is less than the box size but greater than that for the vacuum-wall forming field (which is just the grid spacing). We will set $\beta = \beta_0$ and the net charge $Q=0$ but with a local charge density $\qsubrm{\rho}{Q}\neq 0$. The initial conditions have alternating square regions of charge, each region being $1024$ grid-squares across -- every region has the same magnitude of charge density and adjacent regions being opposite in value. Thus, along a particular horizontal or vertical line, this amounts to the charge density being $+-+-$. Images of the charge density in these simulations can be found in  \fref{fig:sec3:4096_kv_ccor_rho} for two initial maximum charge densities. Naively, one may think that charge will find anti-charge and annihilate. Initially, some annihilation does occur on the boundaries between regions with different charge. However, we find that once charge has condensed onto vacuum-walls it stays there to create a superconducting vacuum-wall. One observes that vacuum-walls have both positive and negative charge condensed on them, as the vacuum-walls pass between regions of different charge densities. The resulting superconducting vacuum-walls do not collapse as readily as their uncharged counterparts -- although the scaling exponent $\gamma$ is not as far from unity in this correlated-charge case as in the homogeneous charge case: $\qsubrm{(1-\gamma)}{cor}<\qsubrm{(1-\gamma)}{homo}$. This is due to some charge density annihilating reducing the amount of charge available to condense onto the walls. 

\section{Generalized Phase Mixing and Separation}
\label{sec:phase-mixing-theory}
In the $U(1)\times\mathbb{Z}_2$ model the parameter $\beta=\beta_0$, the value of the cross coupling above which one is in the phase separation regime, corresponds to the determinant of the quartic coupling matrix being zero. The concept of this ``mixture stability'' occurs in a number of physical systems. The simplest example is that of oil and water where bubbles of oil separate out from the water and are stable to decay essentially due to the system being in the phase separation regime. It has also been observed in two component Bose-Einstein condensates, for example, between the hyperfine states of Rubidium-87 \cite{myatt_bec_sep_exp, stenger_inouye_etal_bec_exp_2}, and can be useful in the creation of topological defects \cite{BattyeCooperSut}. In this section we will derive the equivalent condition for the CCCA model which is a little more complicated due to the additional fourth order terms proportional to $\epsilon$. This will allow us to identify the value of $\beta_0$ in these models and test the concept in subsequent sections.

We derive the mixture stability criterion by computing conditions on the coefficients such that the stable configuration can have both components in mutual existence. To begin, let us write the potential term using (\ref{eq:pot_1b_mca}) as
 \bea
 \label{eq:pot-prenalpha}
V=\half{g_{11}}{}|\Phi|^4 + \half{g_{22}}{}|\sigma|^4  + g_{12}|\Phi|^2|\sigma|^2- m_1|\Phi|^2 - m_2|\sigma|^2 + \epsilon\sum_{i=1}^N\phi_i^4 +V_0,
 \eea
 where we have introduced interaction coefficients $g_{ij}$ and masses $m_i$; this is to make contact with the notation used in the condensed matter literature. The $g_{ij}, m_i$ and $V_0$ are given in terms of the model parameters by
\begin{subequations}
\label{eq:back-terms}
\bea
&g_{11} = \half \lambda_{\Phi},\quad g_{22} = \half \lambda_{\sigma},\quad g_{12} = \beta,& \\
&\quad m_1 = \half \lambda_{\Phi}\eta_{\Phi}^2,\quad m_2 = \half \lambda_{\sigma}\eta_{\sigma}^2,\quad V_0 = \frac{1}{4}\left( \lambda_{\Phi}\eta_{\Phi}^4 + \lambda_{\sigma}\eta_{\sigma}^4\right).&
\eea
 \end{subequations}
In addition, we introduce the number density parameters,
\bea
n_{\Phi} \defn \sum_{i=1}^Nn_i,\quad n_i \defn \phi_i^2,\quad n_{\sigma} \defn |\sigma|^2,
\eea
so that the potential (\ref{eq:pot-prenalpha}) is given by
\bea
V = \half g_{11} n_{\Phi}^2 + \half g_{22} n_{\sigma}^2 + g_{12}n_{\Phi}n_{\sigma} - m_1n_{\Phi} - m_2 n_{\sigma} + \epsilon \sum_{i=1}^Nn_i^2 + V_0.
\eea
To find the field configuration that extremizes the potential, one computes the first derivatives of the potential with respect to each of the density parameters (each $\phi_i$-component to be computed separately) setting them to zero. This implies that
\bea
\label{eq:dvdni=0}
\pd{V}{n_i} &=& g_{11}n_{\Phi} + g_{12}n_{\sigma} - m_1 + 2\epsilon n_i =0,\\
\pd{V}{n_{\sigma}} &=& g_{22} n_{\sigma} + g_{12} n_{\Phi} - m_2 = 0.
\eea
As each of the $N$ components of (\ref{eq:dvdni=0})  are individually zero, so is their sum. Hence, summing (\ref{eq:dvdni=0}) over $i = 1, \ldots, N$ gives
\bea
\pd{V}{n_{\Phi}} \defn \sum_i \pd{V}{n_i}=(g_{11}n_{\Phi} + g_{12}n_{\sigma} - m_1)N + 2\epsilon n_{\Phi} =0.
\eea
This step of constructing an equation in terms of $n_{\Phi}$ and $n_{\sigma}$ without reference to the individual $n_i$ is crucial in what follows -- not all models can be written in such a way, for example, if the anisotropy term has non-diagonal quartic terms.

\begin{figure*}[!h]
      \begin{center}
	\includegraphics[scale=0.7]{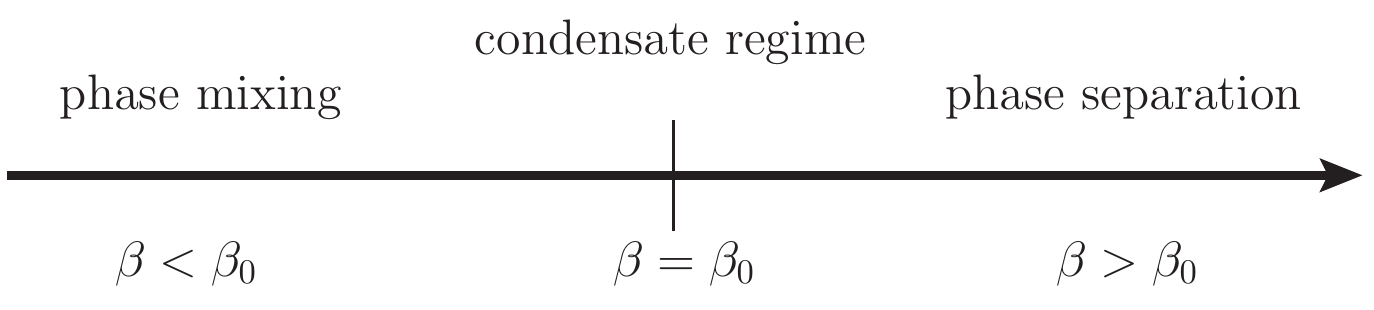}
      \end{center}
\caption{Schematic depicting the parameter ranges for the three phase-interaction regimes of the $U(1) \times \mathbb{Z}_2$ model.}\label{fig:sec3:schematic-psep}
\end{figure*}

The condition that a given configuration is stable comes from the requirement that the eigenvalues of the Hessian are positive-definite. That is, solving
\bea
\left|\pd{^2V}{n_{\alpha}\partial n_{\beta}} - \Lambda \delta_{\alpha\beta} \right|=0\quad\mbox{with} \quad n_{\alpha} \in \{n_{\Phi},n_{\sigma}\},
\eea
and requiring that $\Lambda >0$. This results in the condition
\bea
\label{eq:psep-cond-1}
(g_{11} + 2\epsilon)g_{22} - g_{12}^2 >0.
\eea
If $\epsilon = 0$ the left hand side of the above inequality is simply the determinant of the matrix of quartic coupling coefficients between the components of the mixture. Finally, (\ref{eq:psep-cond-1}) in terms of the model parameters (\ref{eq:back-terms}) is given by
\bea
\beta < \beta_0,\quad \beta_0 \defn\half \sqrt{\lambda_{\Phi}\lambda_{\sigma} + 4\epsilon \lambda_{\sigma}}.
\eea
This is the condition that must be satisfied for a composite mixture of $\Phi$ and $\sigma$ to be stable. Hence, for $\beta > \beta_0$, the mixture is unstable, with the components separating. \fref{fig:sec3:schematic-psep} has a schematic describing the regimes  discussed here.

\section{Charge-coupled Cubic Anisotropy (CCCA) Model}
\label{sec:mca_results}
Let us now turn our attention to the CCCA model. There is much the same phenomenology as in the $U(1)\times \mathbb{Z}_2$ model, the main difference being the presence of multiple vacua, and hence the possibility of junctions for $N>1$. Throughout we take $\lambda_{\sigma} = 1$, $ \eta_{\sigma}=\sqrt{3}/2$ and $\epsilon =-0.1$. As with our $U(1)\times \mathbb{Z}_2$ simulations we present evolution of a random distribution of vacua in the $\Phi$-field in a homogeneous charged background for a variety of values of the coupling parameter, $\beta$, and initial charge density $\qsubrm{\rho}{Q}(0)$; we vary $A$ and fix $\omega = 1$. 

\begin{figure*}[!t]
      \begin{center}
{\includegraphics[scale=0.9]{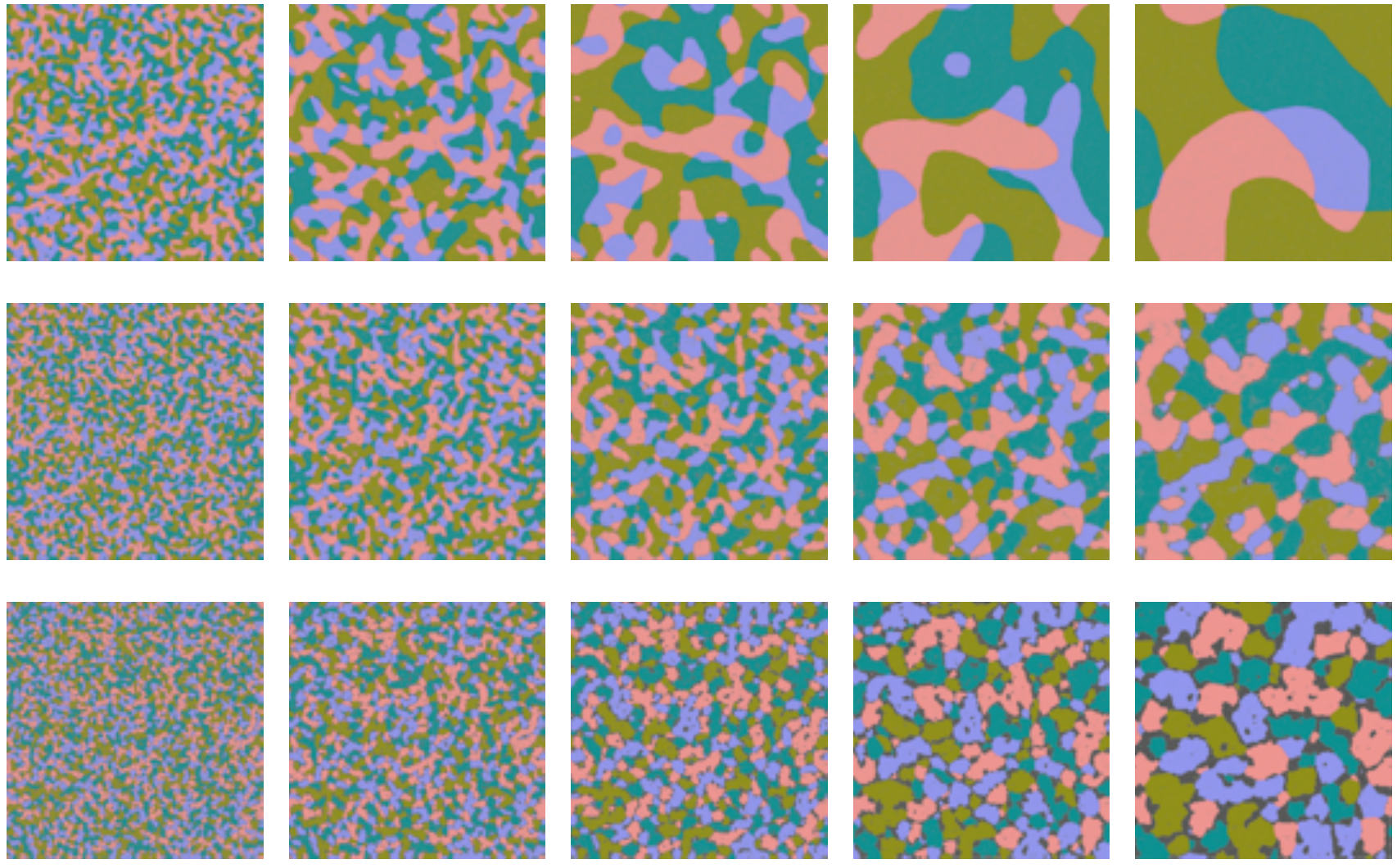}}
\end{center}
\caption{Evolution of the $(\phi_1,\phi_2)$-fields in the regimes $\beta = \half\beta_0, \beta_0, \frac{3}{2}\beta_0$ (top to bottom), at times $t = 800, 1600, 3200, 6400,12800$ (left to right), where $\qsubrm{\rho}{Q}(0)=0.25$ and $P=4096$. Each colour/shade represents one of the four points on the vacuum manifold with black/darkest being $|\Phi|\approx 0$ which corresponds to regions of high charge density.}\label{fig:sec5:o2u1-phasesep}
\end {figure*}

To begin, we vary $\beta$ either side of $\beta_{0}$ as we did for the $U(1) \times \mathbb{Z}_2$ model and take $N=2$. \fref{fig:sec5:o2u1-phasesep} shows the evolution of the $(\phi_1,\phi_2)$-fields where   $\beta = \half\beta_0,\beta_0,\frac{3}{2}\beta_0$. The case of $\beta = \half\beta_0$ is very similar to that in the $U(1)\times \mathbb{Z}_2$ model: vacuum-walls evolve with scaling exponent $\gamma\approx1$ and the background of charge remains approximately homogeneous which is indicative of the phase-mixing regime. The condensate regime $\beta = \beta_0$ clearly does not evolve as quickly, with charge and current becoming associated with the vacuum-walls which appears to inhibit their collapse. It is interesting to note the existence of both $X$- and $Y$-type junctions, especially at late time (i.e. $t = 12800$). The final case $\beta = \frac{3}{2}\beta_0$ is qualitatively very similar to that in the $U(1)\times \mathbb{Z}_2$ model: it appears that the system is in the phase separation regime and we observe component-walls between regions with charge and those without. 

\begin{figure}[!t]
      \begin{center}
      \vspace{1.2cm}
{\includegraphics[scale=0.9]{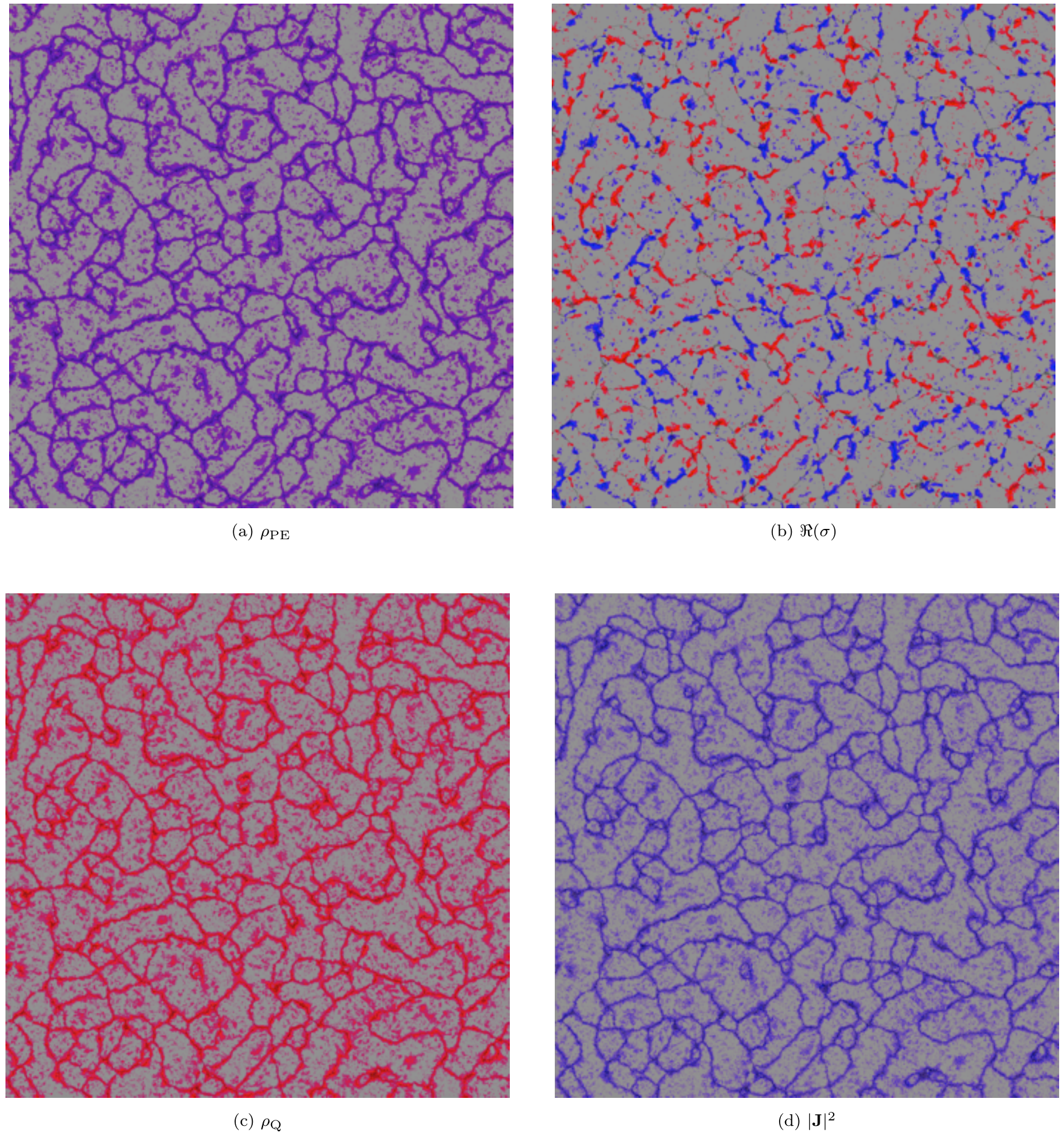}}
\end{center}
\caption{Properties of the field configurations at $t=6400$ for $\beta = \beta_0$ with $N=2,P=4096, \qsubrm{\rho}{Q}(0)=0.25$. The colouring scheme is such that red/blue denotes positive/negative and grey being $\Re(\sigma) < 40\%$ maximal value; $\qsubrm{\rho}{PE,Q} < 10\%$ maximum; $|\rbm{J}|^2 < 2\%$ maximum. These images show that charge and current have become almost exclusively associated with the vacuum-walls. }
\label{fig:images-flds-resigma-largemontage}
\end {figure}

\begin{figure}[!t]
      \begin{center}
            \vspace{1.2cm}
{\includegraphics[scale=0.9]{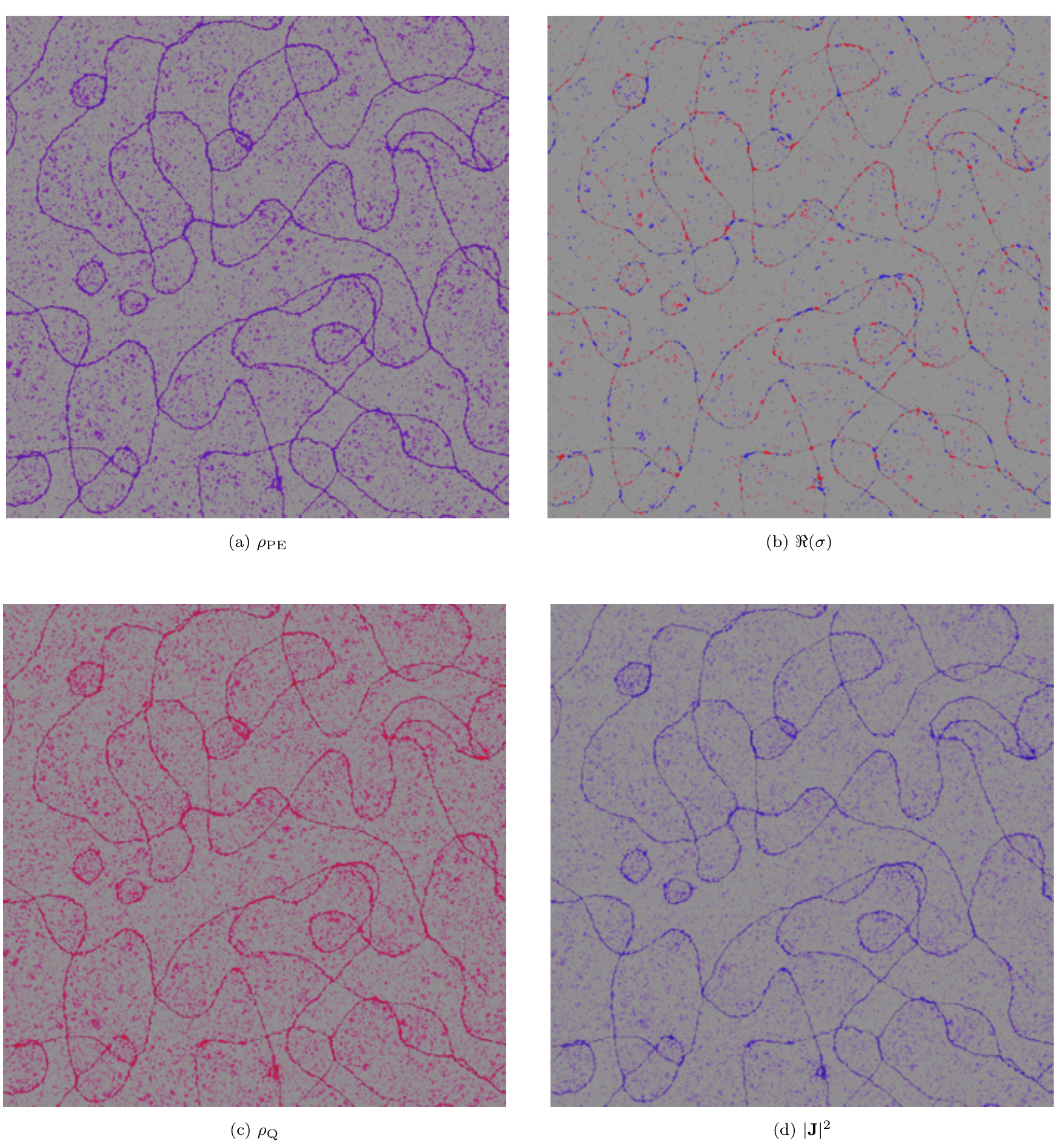}}
\end{center}
\caption{The same as presented in \fref{fig:images-flds-resigma-largemontage} but for a lower initial charge density, $\qsubrm{\rho}{Q}(0) = 0.09$. The formation of junctions and the fact that the $\sigma$-field winds around the walls is more obvious at this lower charge density.}
\label{fig:images-flds-resigma-largemontage-lowq}
            \vspace{1.2cm}
\end {figure}

 \fref{fig:images-flds-resigma-largemontage} shows the configurations of the potential energy density $\qsubrm{\rho}{PE}$, real part of the condensate field $\Re(\sigma)$, charge $\qsubrm{\rho}{Q}$ and current $|\rbm{J}|^2$ at $t=6400$ of the $\beta = \beta_0$ simulation. \fref{fig:images-flds-resigma-largemontage}(a) is not surprising: potential energy peaks on vacuum-walls as the field interpolates over maxima of the potential. \fref{fig:images-flds-resigma-largemontage}(b) shows that $\Re(\sigma)$ varies from positive to negative along the walls which indicates that the $\sigma$-field winds along a vacuum-wall -- the alternating red-blue along a given vacuum-wall. From \fref{fig:images-flds-resigma-largemontage}(c) and (d) it is obvious that both charge and current have become almost exclusively associated with the vacuum-walls. The inspection of an equivalent set of simulations presented in \fref{fig:images-flds-resigma-largemontage-lowq} for a lower initial charge density $\qsubrm{\rho}{Q}(0)= 0.09$ show more clearly the winding. It is the association of charge and current with vacuum-walls  which provides the stabilization mechanism of kinky vortons \cite{BattyePearson-kvform} and it appears from \fref{fig:images-flds-resigma-largemontage} and \fref{fig:images-flds-resigma-largemontage-lowq} that we can extend the mechanism to the CCCA model. If we increase the threshold used to create the images of the charge density $\qsubrm{\rho}{Q}$ (so that the charge density at a given location must be a larger fraction of the highest value than that used in \fref{fig:images-flds-resigma-largemontage} and \fref{fig:images-flds-resigma-largemontage-lowq}) it is revealed that junctions carry a higher charge density than vacuum-walls, $\qsubrm{\rho}{Q,junc}>\qsubrm{\rho}{Q,wall}$.

\begin{figure*}[!t]
      \begin{center}
{\includegraphics[scale=1.0]{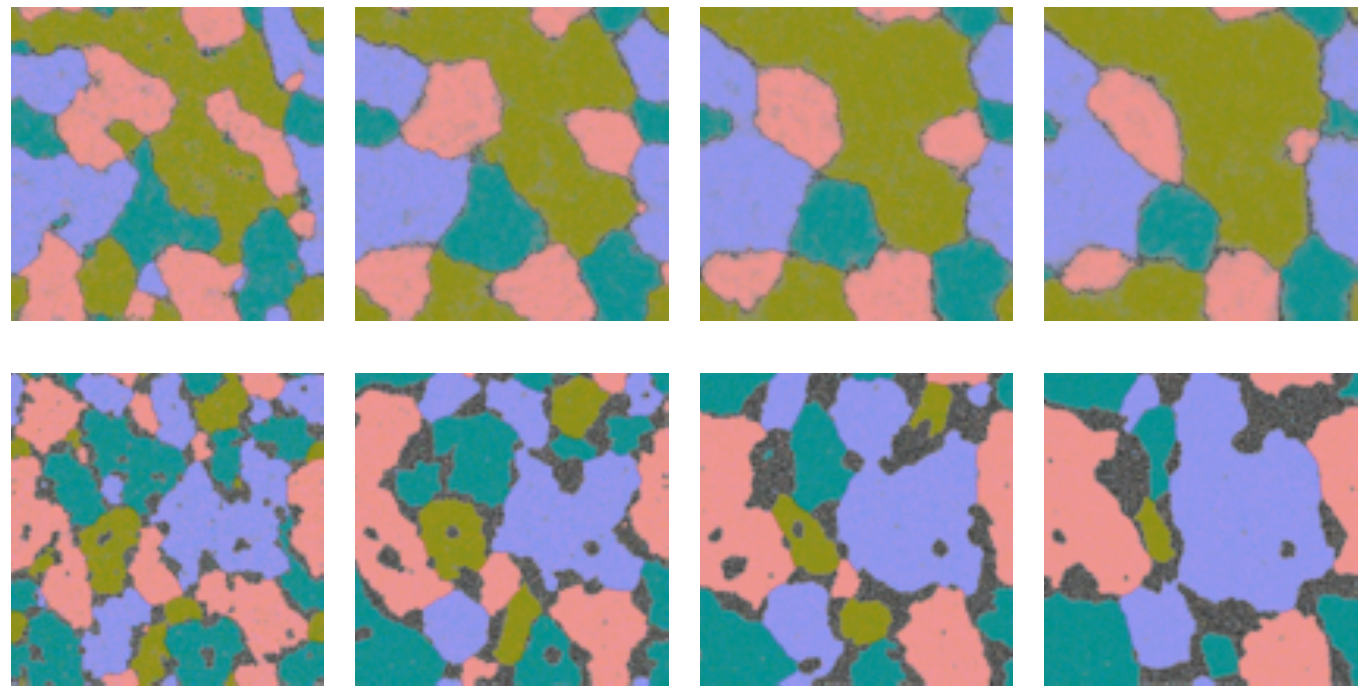}}
\end{center}
\caption{Evolution of the $(\phi_1,\phi_2)$-fields in the regimes $\beta =\beta_0, \frac{3}{2}\beta_0$ (top to bottom), at $t = \tau, 2\tau, 3\tau, 4\tau$ (left to right), where $\qsubrm{\rho}{Q}(0)=0.25$ and $P=1024$. One can observe that structures have persisted beyond the light crossing time $\tau$. In the $\beta = \beta_0$ regime the network's stability is due to symmetry currents generating a superconducting vacuum-wall solution which freezes the network into a glass-like lattice. In the $\beta = \frac{3}{2}\beta_0$ regime phase separation has prevented two adjacent vacuum occupying regions meeting and annihilating (as is the case in the phase mixing or uncharged systems).}\label{fig:sec:o2u1-beta015beta0_taus}
\end {figure*}

The images in \fref{fig:sec:o2u1-beta015beta0_taus} show how the pseudo-stable structures evolve for times beyond the light-crossing time $t \gg \tau$ in the regimes $\beta = \beta_0, \frac{3}{2}\beta_0$ (note that $P = 1024$ was used to reduce computational time). As time progresses in the case of $\beta = \frac{3}{2}\beta_0$ the separation between the vacuum-occupying and charge components becomes more obvious. At late times the configuration with $\beta = \beta_0$ appears to have stabilized into a glass-like state.

\begin{figure*}[!t]
      \begin{center}
{\includegraphics[scale=0.8]{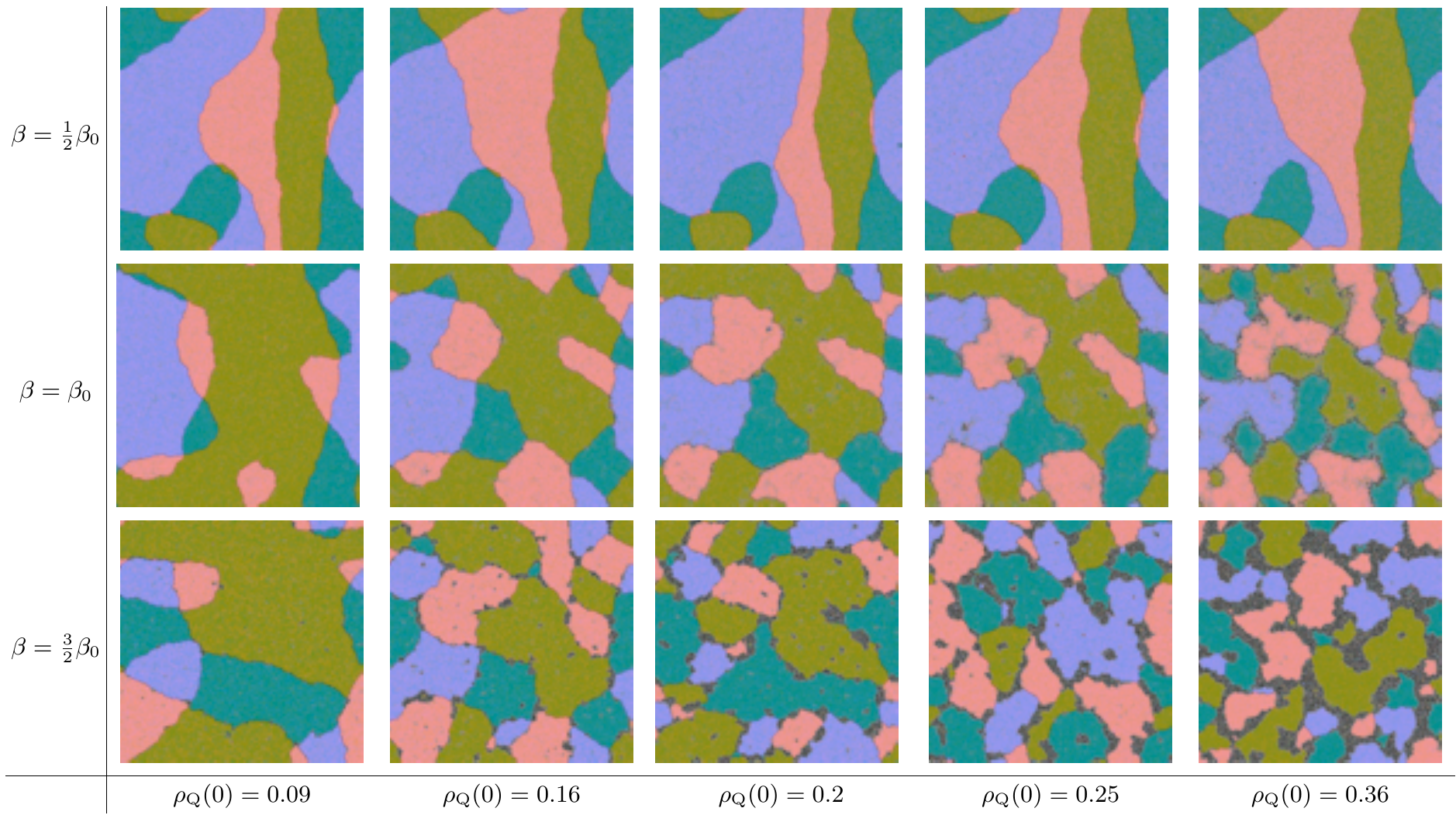}}
\end{center}
\caption{Snap-shots from the evolution of $(\phi_1,\phi_2)$ at $t=3200$, with $P=1024$, in $(\qsubrm{\rho}{Q}(0), \beta)$-space. Note that only vacuum-walls exist for $\beta < \beta_0$ and the existence of component walls for $\beta>\beta_0$. Regardless of the value of $\qsubrm{\rho}{Q}(0)$ in the regime $\beta = \half \beta_0$ the network has collapsing vacuum-walls, with scaling exponent $\gamma \approx 1$. One can observe that as $\qsubrm{\rho}{Q}(0)$ increases in the regime $\beta = \beta_0$, the glass-like nature of the resulting lattice becomes more apparent. As $\qsubrm{\rho}{Q}(0)$ is increased in the $\beta = \frac{3}{2}\beta_0$ regime, there is more charge to separate from the vacuum-occupying regions, which is obvious by noting that the size of the charge-regions increases with initial charge density.}\label{fid:sec:onu1_montage}
\end {figure*}

We have performed  a number of simulations varying the initial charge density $\qsubrm{\rho}{Q}(0)$  and the cross coupling parameter $\beta$. The results  are in \fref{fid:sec:onu1_montage} at $t = 3200$ for $P=1024$. These clarify how the system responds to different amounts of initial charge in the phase separation regime. Interestingly we find that, in the $\beta = \half\beta_0$ simulations, all images are qualitatively identical with $\gamma \approx 1$ irrespective of $\qsubrm{\rho}{Q}(0)$. Finally, as the initial charge density is increased in the condensate regime, it is obvious from the images that the vaccum-walls retain their length for longer.

\begin{figure*}[!t]
      \begin{center}
{\includegraphics[scale=0.9]{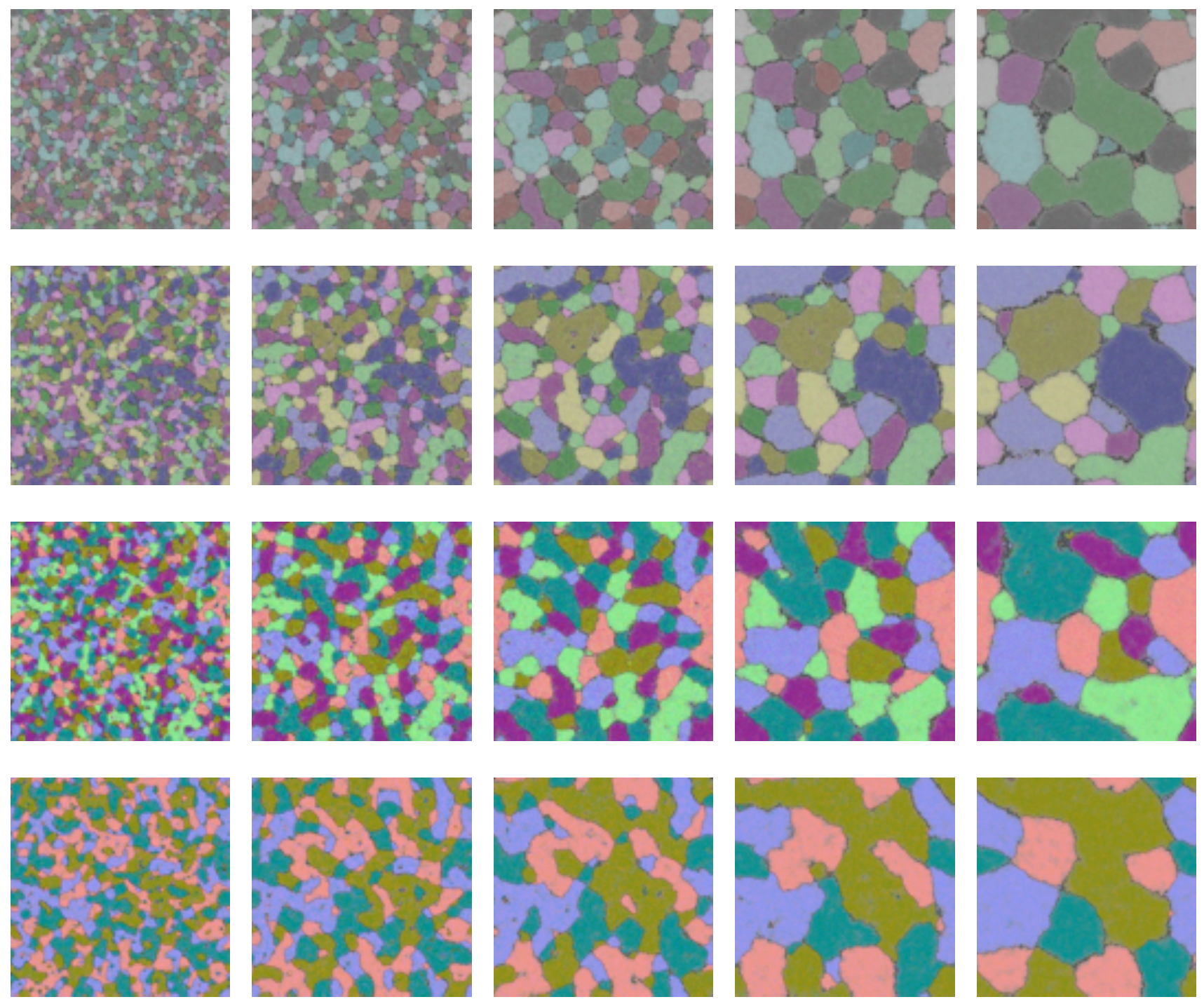}}
\end{center}
\caption{Evolution of the $\phi_i$-fields for $N=5,4,3,2$ (top to bottom), at $t = 400,800,1600,3200,6400$ (left to right), where $\qsubrm{\rho}{Q}(0)=0.25, \beta = \beta_0$ and $P=1024$. As the number of vacua is increased with $N$, the network dynamic is qualitatively identical to the $N=2$ case where charge and current form superconducting vacuum-wall solutions stabilizing the network into a glass-like lattice.}\label{fig:sec:onu1-n=2345}
\end {figure*}    

\begin{figure*}[]
      \begin{center}
      {\includegraphics[scale=1.5]{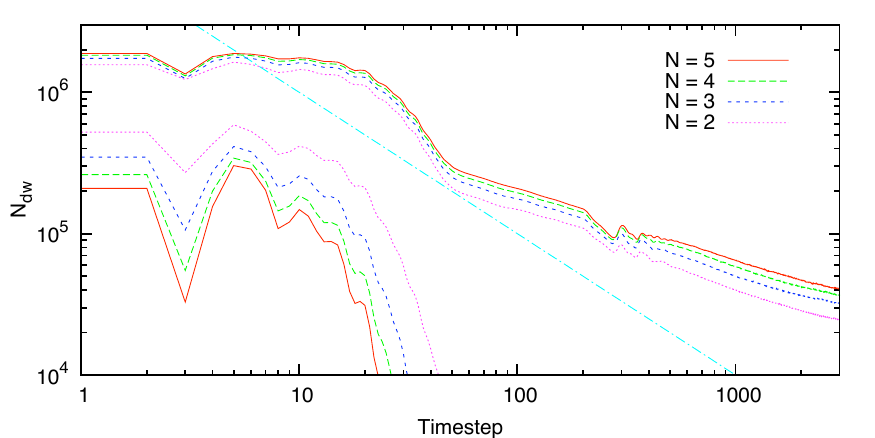}}
      \end{center}
\caption{Evolution of the number of walls in the cubic anisotropy model with $N$ real fields, where $\qsubrm{\rho}{Q}(0)=0.25, P=1024$, for an ensemble average over 10 realisations. The top set of curves are the low tension walls, and the bottom set the high tension walls. One can note that high tension walls decay faster in higher-$N$ models. We have also plotted $\qsubrm{N}{dw} \propto t^{-1}$ for comparison (dotted-dashed line).}\label{eq:sec5:vary-n-flds-onu1}
\end{figure*}

We have also performed a number of simulations for different values of $N$ with $\beta = \beta_0$ and $\qsubrm{\rho}{Q}(0)=0.25$.  Evolution of the fields are presented in \fref{fig:sec:onu1-n=2345} and plots of the evolution of low and high tension vacuum-walls in \fref{eq:sec5:vary-n-flds-onu1}. It appears that the evolution for higher values of $N$ is qualitatively the same as for $N=2$, although the number of high tension walls decays quicker in higher-$N$ models. This is probably due to the larger phase-space of low tension wall trajectories that high tension walls can decay into for large $N$ models. It is clear from the plot of the evolution of the number of walls that the systems have scaling exponents $\gamma < 1$.

\section{Conclusions}
\label{sec:conclusions}

In this paper we have developed the concept of charge condensation onto a random domain wall network and we have given analytic and numerical arguments for its causes and consequences. Furthermore, we have shown that random domain wall systems can form a pseudo-stable glass-like network with the key addition to previous models being the interaction of the field with discrete minima and a field with conserved continuous $U(1)$ symmetry. Because conserved symmetry currents condense on domain walls in a particular parameter regime (what we called the condensate regime) we can extend the mechanism that stabilizes kinky vortons to the rather more general CCCA model. 

The justification for this comes from the observation that in the phase mixing regime the charged field remains homogeneous and does not condense on domain walls, which promptly collapse under their own tension and the network enters a standard scaling dynamic with $\qsubrm{N}{dw} \propto t^{-\gamma}, \gamma = 1$. However, in the phase separation regime the formation of a component-wall appears to prevent vacuum-walls from touching each other and annihilating, thus preventing the natural collapse of wall networks. In the condensate regime we have shown evidence for a winding of the complex scalar field along vacuum-walls and the association of the conserved charge with the wall network. We have shown that the phase separation phenomenon exists in both the $U(1) \times \mathbb{Z}_2$ model and the CCCA model. In both the condensate and phase separation regimes a non-standard scaling behaviour is observed with $\gamma < 1$ and the extent to which a network stabilizes is controlled by the value of the initial charge density. In addition we have also shown that there does not need to be a net charge in a simulation for this stabilization mechanism to occur, merely a local charge density with a sufficiently large correlation length

Controlling whether or not a theory is in the phase separation, phase mixing or condensate regime is achieved by the value of the cross coupling parameter $\beta$, where a given theory has a fixed value $\beta_0$ given in terms of the other quartic coupling parameters. When $\beta > \beta_0$ the theory is in the phase separation regime, when $\beta < \beta_0$ the theory is in the phase mixing regime and when $\beta = \beta_0$ the theory is in the condensate regime.

The dynamics of wall networks was studied extensively in refs. \cite{Avelino:2008ve, Avelino:2006xf, Avelino:2006xy, PinaAvelino:2006ia}. On the basis of these works the authors suggested a ``no frustration conjecture''. They claimed to prove this based on their so called ``ideal model'' which has junctions and walls of identical tensions. The studied scaling regime occurs because the network loses energy as fast as causality will allow; there is nothing to prevent this. The results presented here suggest an important caveat to this: the no frustration conjecture will be true only in the absence of charge, or some other phenomena which prevents the collapse of walls.

We have shown that it may be possible to stabilize networks. Hence, we have provided theoretical evidence which might lead to a physical manifestation of the elastic dark energy model, which, until the charge-coupling mechanism presented here, had been hampered by the $\propto t^{-1}$ scaling behavior of domain wall networks. However. still more works needs to be done before we can claim to have a fully consistent framework for domain walls to provide the dark energy of the Universe.

\section*{Acknowledgements}
We have benefited from code written by Chris Welshman which formed the basis of our visualization software; and from discussions with Paul Sutcliffe and Simon Pike in the early stages of this work. 

\bibliographystyle{apsrev}
\bibliography{references}
\end{document}